\documentclass[aps,pre,twocolumn]{revtex4-1}
\usepackage{amssymb}
\usepackage{amsmath,bm}
\usepackage{graphicx,color}
\usepackage{bbm}
\usepackage[bottom]{footmisc}
\usepackage{multirow}
\newcommand{\B}[1]{{\bm{#1}}}%% Bold Roman & Greek Lower & Upper Case

\begin{document}
\title{Creep failure of amorphous solids under tensile stress}
\author{Bhanu Prasad Bhowmik $^1$, 
	H.G.E. Hentschel$^{1,2}$ and Itamar Procaccia$^{1,3}$}
\affiliation{$^1$Dept. of Chemical Physics, The Weizmann Institute of
	Science, Rehovot 76100, Israel\\
$^2$ Dept. of Physics, Emory University, Atlanta Ga. 30322, USA\\$^3$  Center for OPTical IMagery Analysis and Learning, Northwestern Polytechnical University, Xi'an, 710072 China. }

\begin{abstract}
Applying constant tensile stress to a piece of amorphous solid results in a slow extension, followed by an eventual rapid mechanical collapse. 
This ``creep" process is of paramount engineering concern, and as such was the subject of study in a variety of materials, for more
than a century. Predictive theories for $\tau_w$, the expected time  of collapse, are lacking, mainly due to its dependence on a bewildering
variety of parameters, including temperature, system size, tensile force, but also the detailed microscopic interactions between
constituents.  The complex dependence
of the collapse time on all the parameters is discussed below, using simulations of strip of amorphous material. Different scenarios are observed for ductile and brittle materials, resulting in serious difficulties in creating an all-encompassing theory that could offer safety measures for given conditions. A central aim of this paper is to employ scaling concepts, to achieve data collapse for the {\em probability distribution function} (pdf) of $\ln{\tau_w}$. The scaling ideas result in a universal function which provides a {\em prediction} of the pdf of $\ln{\tau_w}$ for out-of-sample systems, from measurements at other values of these parameters. The predictive power of the scaling theory is demonstrated for both ductile and brittle systems.  Finally, we present a derivation of universal scaling function for brittle materials. The ductile case appears to be due to a plastic necking instability and is left for future research. 
\end{abstract}
\maketitle

\section{Introduction}
The mechanical collapse of a solid under a constant tensile force is known as ``creep failure". Being of a central concern to material physics and engineering, this process has been widely studied over the years in a variety of materials  \cite{AndradeCreep1, AndradeCreep2}.  Creep failure in amorphous solids in particular was examined using experiments, simulation and analytical consideration \cite{BonnBerthierMannevileRMP, NicolasMartensBarrat}. Examining the available data one concludes that creep failure is influenced by a host of variables, causing wide changes in the failure scenarios in different systems. Broadly speaking, one can distinguish ``brittle" and ``ductile" materials. In the former class systems can withstand relatively small tensile forces for a long (experimental) time, but they can collapse immediately in a catastrophic manner for larger forces \cite{BonnScience}. Ductile materials, on the other hand, can exhibit a necking instability, with the system extending its length over a long stretch of time until it finally collapses \cite{03ELP}. But even within these classes, it was found that the mechanism for failure can differ from system to system. Many crucial parameters appear to play an important role, including system size $N$, the system's aspect ratio $A=$length/width, temperature $T$, type of microscopic interactions etc. 

The aim of this paper is to explore the dependence of the creep scenario on all these variables, by using numerical simulations of a simple classical glass former for which all the parameters are under control. Having done so, a second aim is to overcome the complex dependence on the parameters, by finding a method to collapse the data for the probability distribution functions (pdf) of the logarithm of the waiting time to failure ${\ln \tau_w}$. The main result of the paper is that this pdf has a log-normal form that can be collapsed on a universal form, shared by  systems having different parameters. With this at hand, one can determine the universal function from measurements, and then use it to predict this important pdf for out-of-sample systems. Examples of such predictability are discussed below.

The study of creep collapse includes two connected, but nevertheless different, issues. The first is {\em how it occurs} and the second is {\em how long does it take}. We argue here that the former is a very difficult problem, involving a variety of microscopic and macroscopic ingredients. The latter appears manageable, using scaling concepts as shown below. Indeed, the actual scenario for creep failure appears very rich.  At small applied forces the materials extend slowly;  various studies show that the strain $\gamma$ has a power law dependence on time; ${\gamma} \sim t^{\alpha}$ \cite{AndradeCreep1, AndradeCreep2}. The value of the exponent $\alpha$ was estimated in many experiments and simulations for different viscoelastic materials. Pioneering measurements were reported first by Andrade for crystalline metals, with the estimate  $\alpha \approx  \frac{1}{3}$ \cite{AndradeCreep1, AndradeCreep2}. Later studies of various disordered viscoelastic materials yield different values of $\alpha$ in the range [0,1]. In Ref.~\cite{DBAndManneVille2011} this range of values was attributed to the microscopic different origins of the creep flow. In ref. \cite{PinakiPrecursorCreep}
numerical simulations revealed fluctuations in the rate of deformation, identified as precursors to the final rupture or fluidization. 

On the face of it, also the second issue of the waiting time-to-failure $\tau_w$ seems complex. Obviously, $\tau_w$ depends on the applied tensile force. But the nature of this dependence changes drastically in different conditions and it depends on  microscopic interactions. Athermal systems exhibit a power law dependence on the applied force, but at finite temperatures, where the creep is caused by thermal fluctuations, one finds an exponential decrease in the time for failure with the increase in force \cite{Skrzeszewska2010, DBAndManneVille2011, GFAndManneville2010, SBV2012, PRLTheory}. Moreover, the parameters in the exponential dependence vary widely from system to system. Nevertheless we will argue below that the statistics of the time to failure are easier to control and predict than the mode of failure.  

In Sect.~\ref{simulations} we introduce the model employed in the rest of the paper. We choose a ternary mixture of point-particles interacting via a modified Lennard-Jones potential. By varying the range of interaction of the microscopic potential we can produce more brittle or more ductile amorphous material and cf. Ref.~\cite{11DKPZ}.
Indeed, the degree of brittleness is important in determining very different scenarios for creep failure. In Sect.~\ref{observation} we present the qualitative observations, making clear distinction between the mode of failure of brittle and ductile systems. In Sect.~\ref{quant} we turn to quantitative measurements, determining the dependence of the time of collapse on the various parameters, like tensile force  $\B F_{app}$, system size $N$, aspect ratio $A$ and temperature $T$ for both ductile and brittle configurations. 
In all our simulations the final rupture does not occur immediately after the external force is applied;  rupture takes time,  and we refer to this as the waiting time $\tau_w$. We measure the displacement of the center of mass of the strip in the direction of  $\B F_{app}$ which determines the stain ($\gamma$) as a function of simulation time ($t$). We systematically tune the temperature, aspect ratio, number of particles and degree of ductility of the system and observe how the waiting time varies. Since our systems are thermal, $\tau_w$ has exponential dependence on   $\B F_{app}$. The parameters in this relation depend on the temperature, aspect ratio, degree of brittleness and number of particles of the system. The important result of this section is that although the time for collapse $\tau_w$ exhibits large sample-to-sample fluctuations, indeed varying over many orders of magnitude, the {\em average} over realizations having the same parameters, of the logarithm of the waiting time $\langle \ln \tau_w\rangle$, satisfies very simple an reproducible dependence on the various parameters. This will form the basis
of the predictability of the statistics of the time for failure, which is discussed and explained in 
Sect.~\ref{stat}. In that section we show that the pdf of the waiting time assumes a simple log-normal form that can be rescaled to provide a universal pdf that is the basis for the of out-of-sample predictions. In Sect.~\ref{theory} we present a theory for the brittle case, to explain the origin of the log-normal distribution. The ductile case appears to be due to a different mechanism, namely a plastic necking instability \cite{03ELP}, and so we leave it for future studies. In Sect.~\ref{summary} we offer a summary and a discussion of the new results.

\section{Simulations}
\label{simulations}
To allow us to span brittle to ductile behavior we employ a two dimensional amorphous strip  of length $L$ and width $W$ made of a ternary mixture of Lenard-Jones particles, denoted as A, B and C, with a concentration ratio A:B:C =  54:29:17. The particles interact via a modified Lennard-Jones potential-
%\begin{eqnarray}
%&&	V_{\alpha,\beta}(r) = 4\epsilon_{\alpha\beta} \Big[\left(\frac{\sigma_{\alpha\beta}}{r}\right)^{12} - \left(\frac{\sigma_{\alpha\beta}}{r}\right)^{6}
%	+C_0 \nonumber+ C_2\left(\frac{r}{\sigma_{\alpha\beta}}\right)^{2} \nonumber\\&&+  C_4\left(\frac{r}{\sigma_{\alpha\beta}}\right)^{4}\Big] \ , 
%	\label{LJ}
%\end{eqnarray}
\begin{eqnarray}
&&V_{\alpha,\beta}(r) =
4\epsilon_{\alpha\beta} \Big[\left(\frac{\sigma_{\alpha\beta}}{r}\right)^{12} - \left(\frac{\sigma_{\alpha\beta}}{r}\right)^{6}\Big]\quad  \text{if} \left(\frac{r_{ij}}{\sigma_{ij}}\right) \leq r_{min} \nonumber\\
&&V_{\alpha,\beta}(r) =\epsilon_{\alpha\beta} \Big[a\left(\frac{\sigma_{\alpha\beta}}{r}\right)^{12} - b\left(\frac{\sigma_{\alpha\beta}}{r}\right)^{6}
	+C_0 \nonumber+ C_2\left(\frac{r}{\sigma_{\alpha\beta}}\right)^{2} \\  &&+ C_4\left(\frac{r}{\sigma_{\alpha\beta}}\right)^{4}\Big] \quad \text{if}  \quad r_{min} < \left(\frac{r_{ij}}{\sigma_{ij}}\right)\leq r_{c} \nonumber \\	
&&V_{\alpha,\beta}(r) =0 \quad \text{if} \left(\frac{r_{ij}}{\sigma_{ij}}\right) > r_c \ .
 \label{LJ}
\end{eqnarray}

\begin{figure}[ht]
\includegraphics[scale = 0.55]{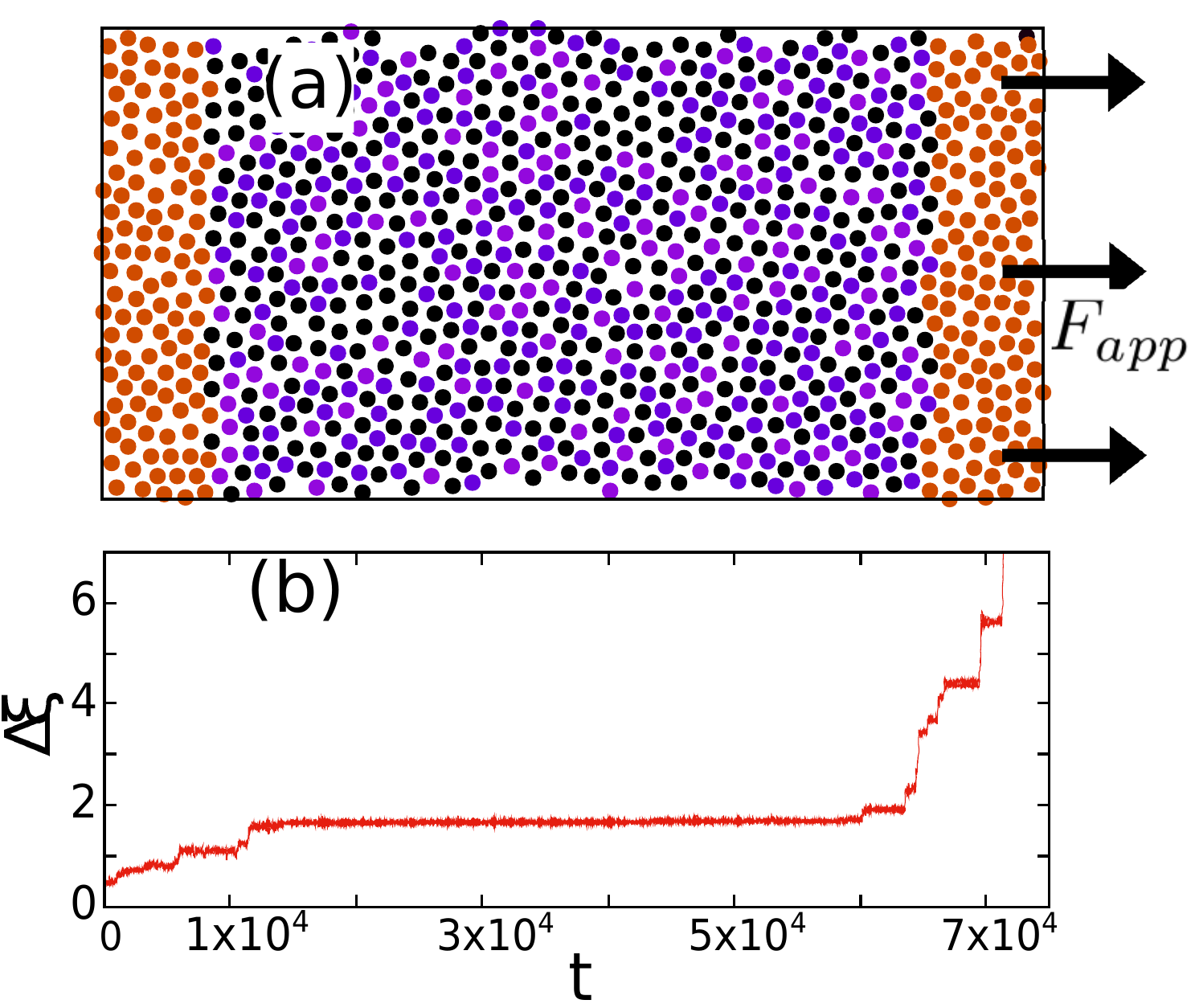}
\caption{Panel (a): A typical configuration of the strip for $N = 1000$. The orange particles consist the walls, and the arrows indicate the direction of applied force. Panel (b): time dependence of the change of center of mass, $\Delta \xi(t)$. Note that time is measured in units of  $\tau = \sqrt{m\sigma_{AA}^2/\epsilon_{AA}}=1$, while in the waiting regime $\Delta \xi(t)\sim 2$ in this simulation.}
\label{fig1}
\end{figure}

Here $\alpha$ and $\beta$ stand for different types of particles. The potential has a minimum at $r = r_{min}\sigma_{\alpha,\beta}$ and it vanishes at $r = r_c\sigma_{\alpha,\beta}$. The value of $r_{min}$ is given by $2^\frac{1}{6}$ and $r_c$ is varied from 1.2 to 2.5 in order to tune the ductility of the system \cite{11DKPZ}.  The coefficients a, b, $C_0$, $C_2$ and $C_4$ are chosen such that the repulsive and attractive parts of the potential are
continuous with one derivatives at the potential minimum, $r_{min}\sigma_{\alpha\beta}$, and the potential goes to zero continuously at $r_c\sigma_{\alpha,\beta}$ with two continuous derivatives. The energy scales are $\epsilon_{AB}$ = $1.5\epsilon_{AA}$, $\epsilon_{BB}$ = $0.5\epsilon_{AA}$, $\epsilon_{AC}$ = $0.5(\epsilon_{AA} + \epsilon_{AB})$, $\epsilon_{BC}$ = $0.5(\epsilon_{AB} + \epsilon_{BB})$ and $\epsilon_{CC}$ = $0.5(\epsilon_{AA} + \epsilon_{BB})$, with $\epsilon_{AA}$ equal 1.
The ranges of interaction are $\sigma_{AB}$ = $0.8\sigma_{AA}$, $\sigma_{BB}$ = $0.88\sigma_{AA}$, $\sigma_{AC}$ = $0.5(\sigma_{AA} + \sigma_{AB})$, $\sigma_{BC}$ = $0.5(\sigma_{AB} + \sigma_{BB})$ and $\sigma_{CC}$ = $0.5(\sigma_{AA} + \sigma_{BB})$, with $\sigma_{AA}$=1. The mass $m$ of all the particles is unity, and the unit of time is $\tau = \sqrt{m\sigma_{AA}^2/\epsilon_{AA}}=1$. Boltzmann's constant is taken as unity. The strip has walls of suitable thickness at the two lateral sides, where the tensile force is applied. The top and bottom sides have open boundaries as shown in Fig.~\ref{fig1}. The simulations are performed at three different temperatures, $T = 0.06, 0.10, 0.15$, for three different system sizes, for which the number of particles $N =$ 1000, 4000 and  10000. In addition, the aspect ratio of the strip, $A \equiv L/W$, is varied from $1.25$ to $3.0$. Note that the simulations are carried out at finite temperatures, where mechanical failure is expected to be sensitive to thermal fluctuations \cite{BHP2022EPL, BHP2022PRE}.
%%%%%%%%%%%%%%%%%%%%%%%%%%%%%%%%%%%%%%%%%%%%%%%%%
%%%%%%%%%%%%%%%%%%%%%%%%%%%%%%%%%%%%%%%%%%%%%%%%%
\begin{figure*}
	\includegraphics[scale = 0.60]{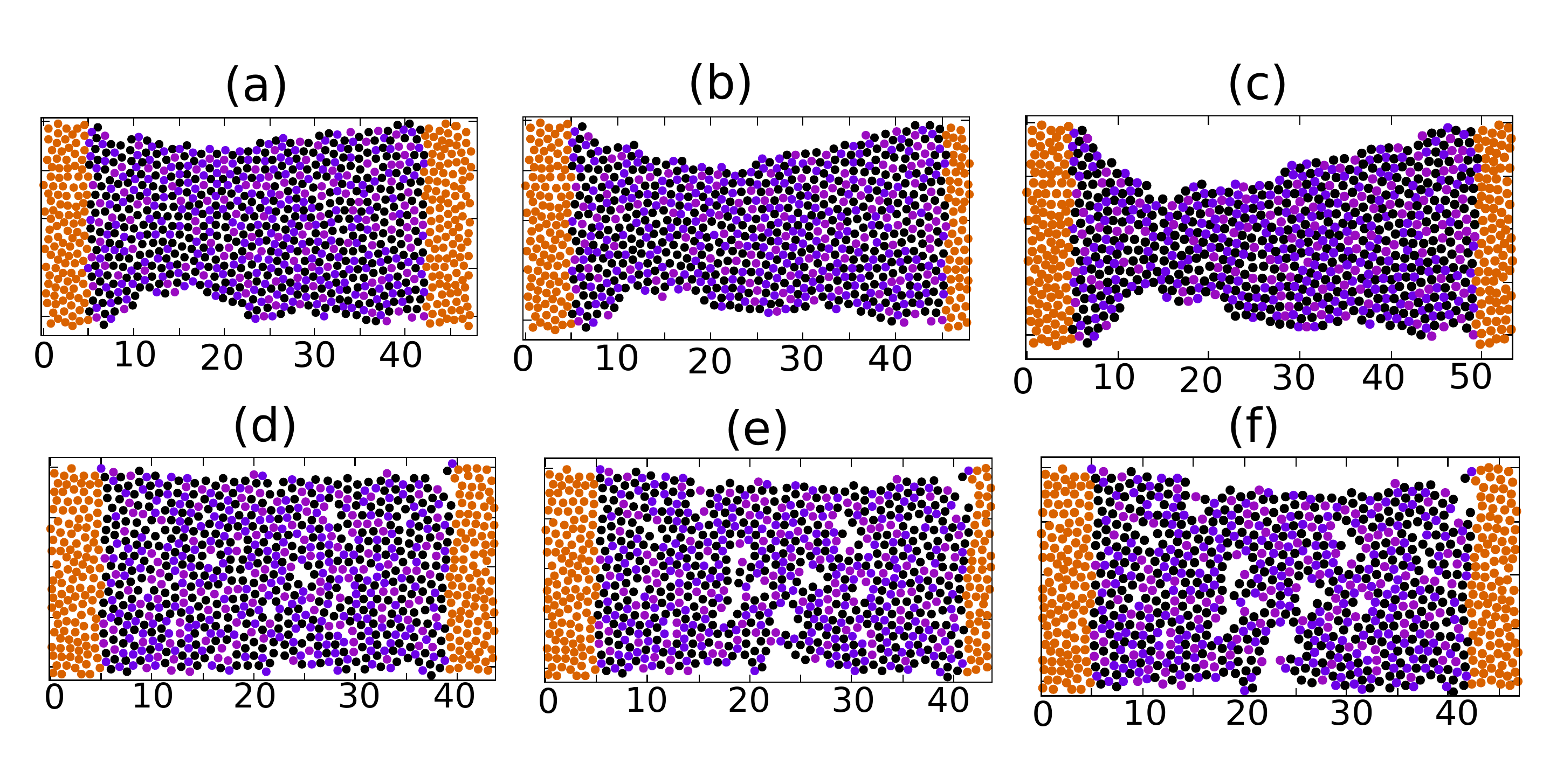}
	\caption{Typical configurations at different simulation time with a constant applied force $\B F_{app}$. Panels a,b,c are for a ductile system and panels d,e,f are for a brittle system.}
	\label{fig2}
\end{figure*}

\begin{figure}
	\includegraphics[scale = 0.7]{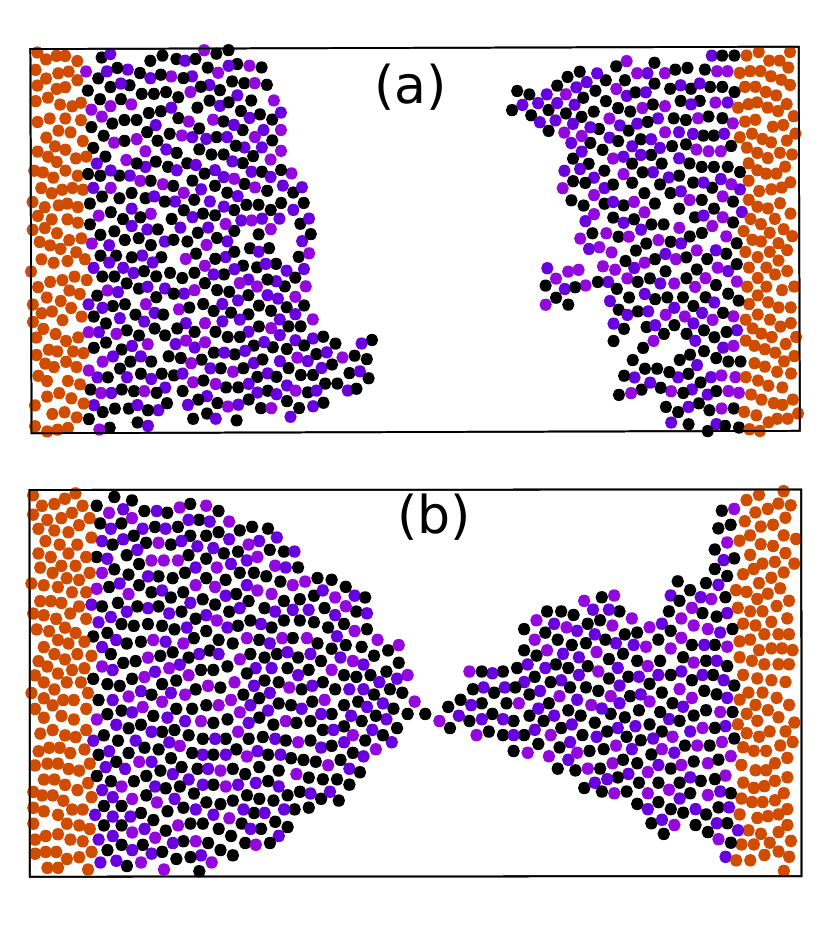}
	\caption{Comparison between the shapes of the interfaces after brittle or ductile failures. Panel (a): typical shape of the interface of a brittle
		strip whose length is of the order of  $W$. Panel (b): typical ductile interface whose length exceeds $W$ due to the necking instability.}
	\label{fig3}
\end{figure} 

All the simulations commence by creating a strip of amorphous matter with periodic boundary condition (PBC) in the $x$ and $y$ directions, where the $x$ direction is determined by the longer boundary, which is also the direction of the tensile force. The strip is equilibrated at a high temperature $T = 1$. Subsequently the system is slowly cooled down to a target temperature $T$. Upon reaching the target temperature we conduct NPT molecular dynamics at zero pressure, $P = 0$. The purpose of this  is to avoid particles escaping from the simulation cell after the removal of the periodic boundary conditions. Upon removing the PBC in the $y$ direction, we construct two boundary walls at the opposite sides in the $x$ direction. The walls have a thickness of around five particle diameters. In the $y$ direction the strip has open boundaries. Finally, we apply a net external force $\B F_{app}$ on one of the walls by equally dividing it among the wall particles. The particles of the other wall remain pinned at their initial position. We continue our simulation until the strip fails mechanically. The force is increased from zero to the chosen value of $F_{app}$ smoothly, to avoid any possible shock. Note that the time required to reach the selected force is very small compared to the typical failure time, $\tau_w$, even for the largest $\B F_{app}$. The failure event is signaled by the relatively rapid increase of the center of mass of the strip in the direction of $\B F_{app}$, cf. Fig.~\ref{fig1}.

\section{Qualitative Observations}
\label{observation}

In this section we explore the modes of mechanical failure for both brittle and ductile systems. Switching on $\B F_{app}$, one can monitor the increase in the position of the center of mass of the strip.
The presence of the tensile force $\B F_{app}$ results, even at short times, with some amount of stretching of the strip, but no failure is observed for small amounts of stretching. There exists an amount of stretching that is admissible without breaking bonds, both in brittle and in ductile materials. Exceeding this amount, bonds begin to break. Typically,  more ductile system stretch more than brittle systems before failing, cf. Fig.~\ref{fig2}. With all the parameters (besides the microscopic interaction law) kept equal, brittle systems need larger values of 
$\B F_{app}$ to fail, while increasing the temperature $T$ results in smaller values of $\B F_{app}$
in order to fail. The required  $\B F_{app}$ also increases upon increasing the  aspect ratio $A$ of the strip, as the number of broken bonds leading to final rupture increases. For ductile failure the system fails by forming a neck, the bonds at the edges start breaking, resulting in an increase in the applied force per bond in the remaining bonds. On the other hand, in brittle systems there are multiple holes in the bulk of the strip, with occasional micro-cracks at the edges \cite{21DPH}. These holes and cracks can merge, leading to the final rupture, as shown in Fig.~\ref{fig2}. Again,  the formation of the holes and cracks  necessarily increases the applied force per remaining bond. Thus both for brittle and ductile samples the increase in tensile force per bond initiates the process of mechanical failure, but the details of this process depend crucially on the microscopic interaction. The appearance of the strip after failure for ductile and brittle systems are shown in Fig.~\ref{fig3}. 

Below we will make use of the hole formation in the brittle scenario to derive the pdf of $\ln\tau_w $. The observation is that the holes are created on a relatively fast time scale, 
and then $\tau_w$ is determined by the additional bonds that need to be broken by thermal fluctuations. The number of bonds is proportional to the sum of circumferences of the holes.
Since the number and size of the holes that are created on the fast time scale is random, 
the distribution of the total circumference is a normal (Gaussian) distribution. Adding to this the fact that the thermal fluctuation needed to overcome the barrier for breaking bonds follows a Kramer's theory \cite{40Kra}, we will be able to relate this geometric picture to the pdf of $\ln\tau_w$ as is explained in Sect.~\ref{theory}. Unfortunately we do not have such a simplifying geometric picture for the ductile scenario where the stretching of the system, and the length of the neck that forms, are harder to estimate using simple considerations. 

\section{Quantitative Observations}
\label{quant}

Our simulations reveal that the time for failure $\tau_w$ depends on all the parameters, i.e.
$N$, $\B F_{app}$, $T$ and $A$,
\begin{equation}
\tau_{w} \sim  \tau_{w}\left(N,F_{app}, T, A \right) \ .
\label{eqn1}
\end{equation}
Moreover, even for the same parameters, different samples can exhibit widely different values of $\tau_w$.
Therefore, to make progress, one needs to consider the mean values, like $\langle \tau_w\rangle$, $\ln \langle \tau_w\rangle$ or $\langle \ln \tau_w\rangle$, where $\langle \cdots\rangle$ represents an average over an ensemble of realization sharing the same parameters
$N$, $\B F_{app}$, $T$ and $A$. All these mean values obey simple relationship to the control parameters as we discuss below. In light of the theoretical consideration presented in Sect.~\ref{theory}, we prefer to detail simulations results for  $\langle \ln \tau_w\rangle$, an average quantity that plays a dominant role in the theory below. 
The figures shown below, i.e. Fig.~\ref{fig4} - \ref{fig7} are all extracted from data
pertaining to $R_c=2.5$. Similar results were found other values of $R_c$, and cf. Fig.~\ref{fig8}.

\subsection{Dependence on tensile force and temperature}

\begin{figure}
\includegraphics[scale = 0.43]{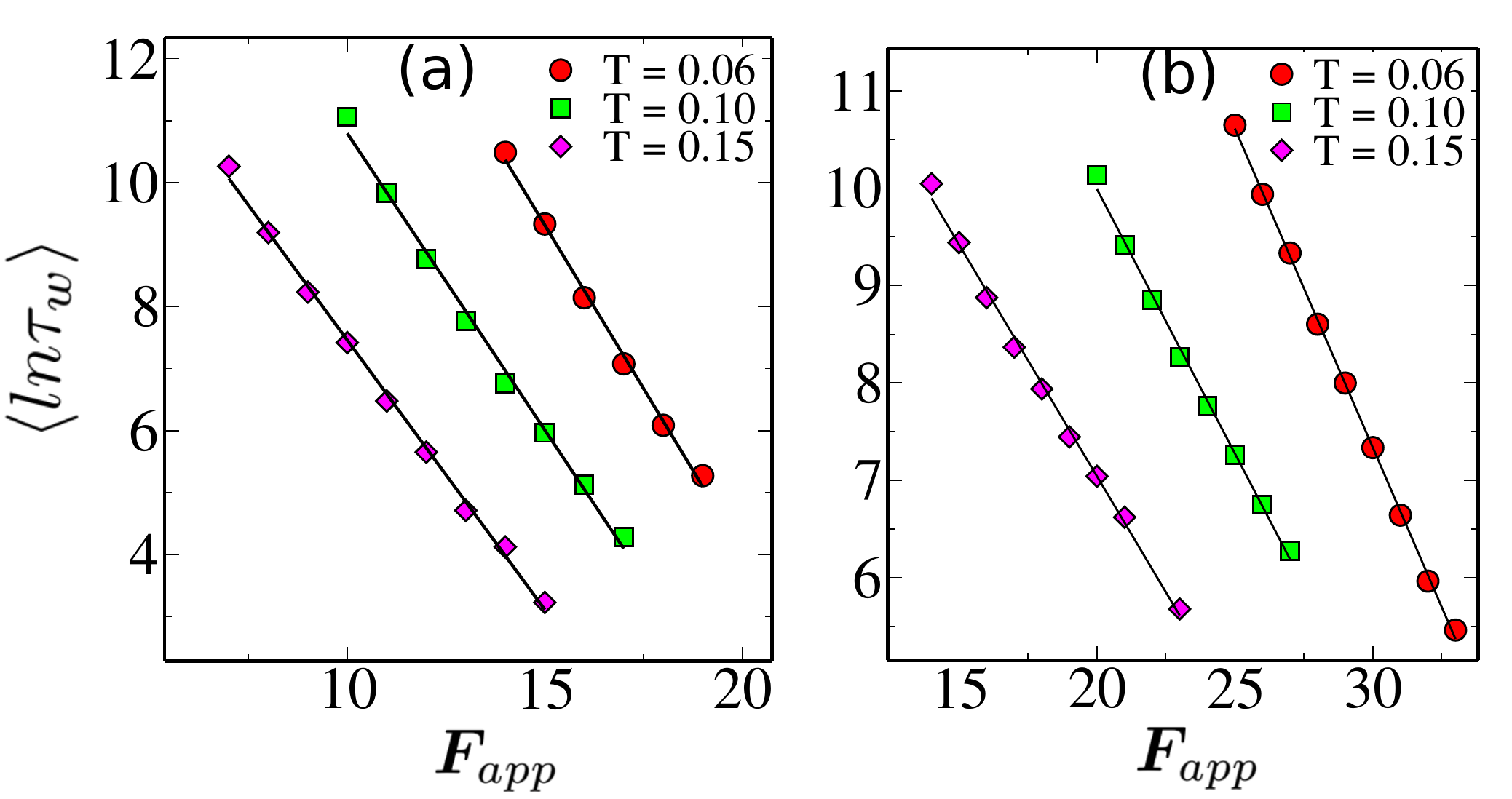}
\caption{Panel (a): Average logarithm of waiting time as a function of applied force for different temperatures, for $N = 1000$. The solid lines are the fits to the Eq. (\ref{eqn2}). Note that the slope, $D_1(T)$ depends on $T$. Panel (b): Same for $N = 4000$.}
\label{fig4}
\end{figure}

\begin{figure}
\includegraphics[scale = 0.4]{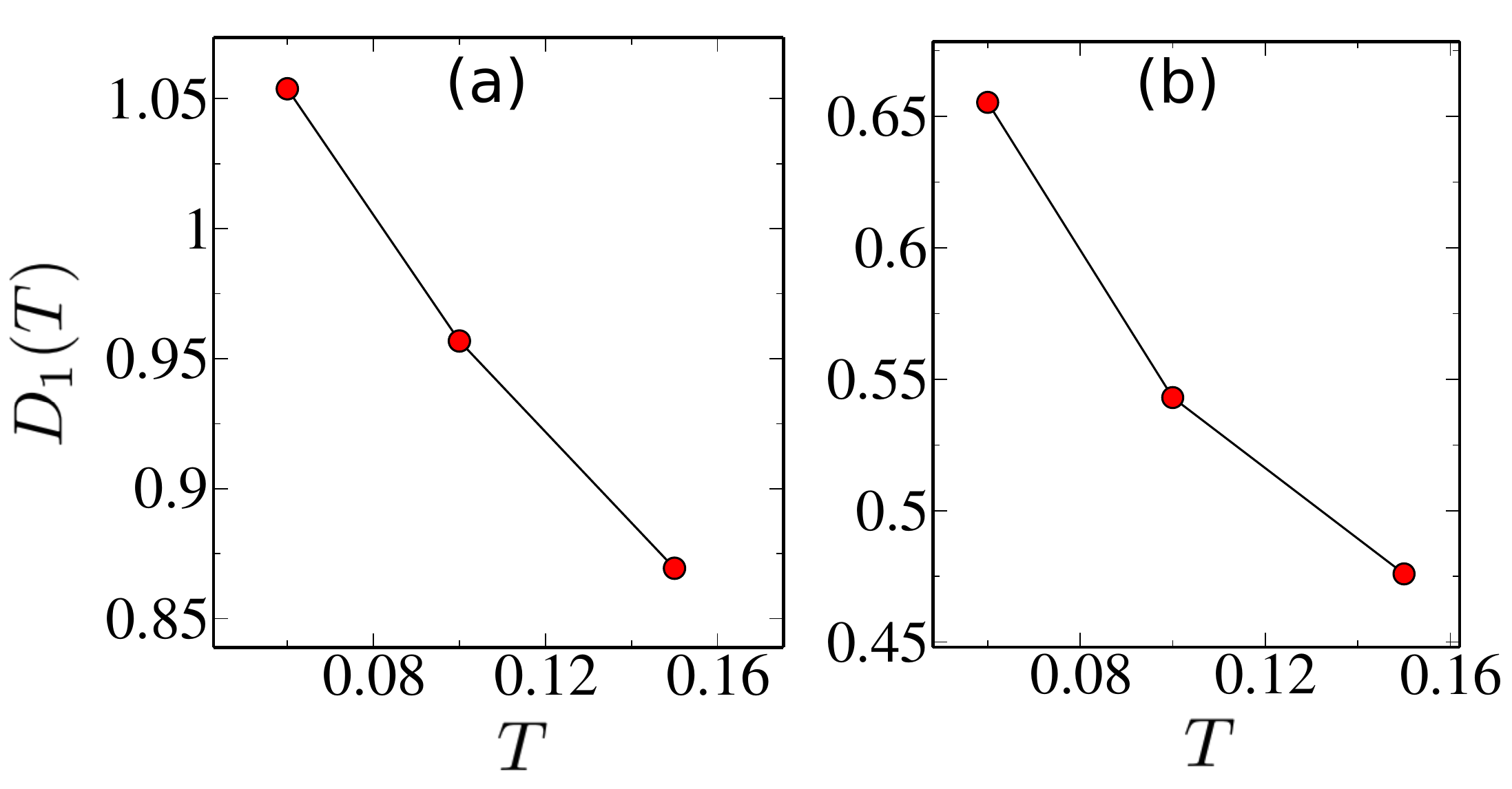}
\caption{Temperature dependence of the $D_1(T)$ for $N = 1000$ (panel (a)) and $N = 4000$ (panel (b)).}
\label{fig5}
\end{figure}

By creating an ensemble of configurations sharing the same aspect ratio and number of particles $A$ and $N$, one finds a simple law for the dependence of the mean time to fracture on the tensile force and temperature:
\begin{equation}
\langle\ln  \tau_{w}\left(F_{app}, T \right)\rangle \approx C_1 - D_1 (T) {F}_{app} \ .
\label{eqn2}
\end{equation}
The demonstration of this simple law for two different system sizes is shown in Fig.~\ref{fig4}. The temperature dependence of the coefficient $D_1(T)$ in Eq.~\ref{eqn2} is shown in Fig.\ref{fig5}.
%%%%%%%%%%%%%%%%%%%%%%%%%%%%%%%%%%%%%%%%%%%%%%%%%%%%%%%%%%%%%%%%%%%
%%%%%%%%%%%%%%%%%%%%%%%%%%%%%%%%%%%%%%%%%%%%%%%%%%%%%%%%%%%%% 
%%%%%%%%%%%%%%%%%%%%%%%%%%%%%%%%%%%%%%%%%%%%%%%%%%%%%%%%%%%%%%%

\subsection{Dependence on aspect ratio $A$}
\begin{figure}
\includegraphics[scale = 0.45]{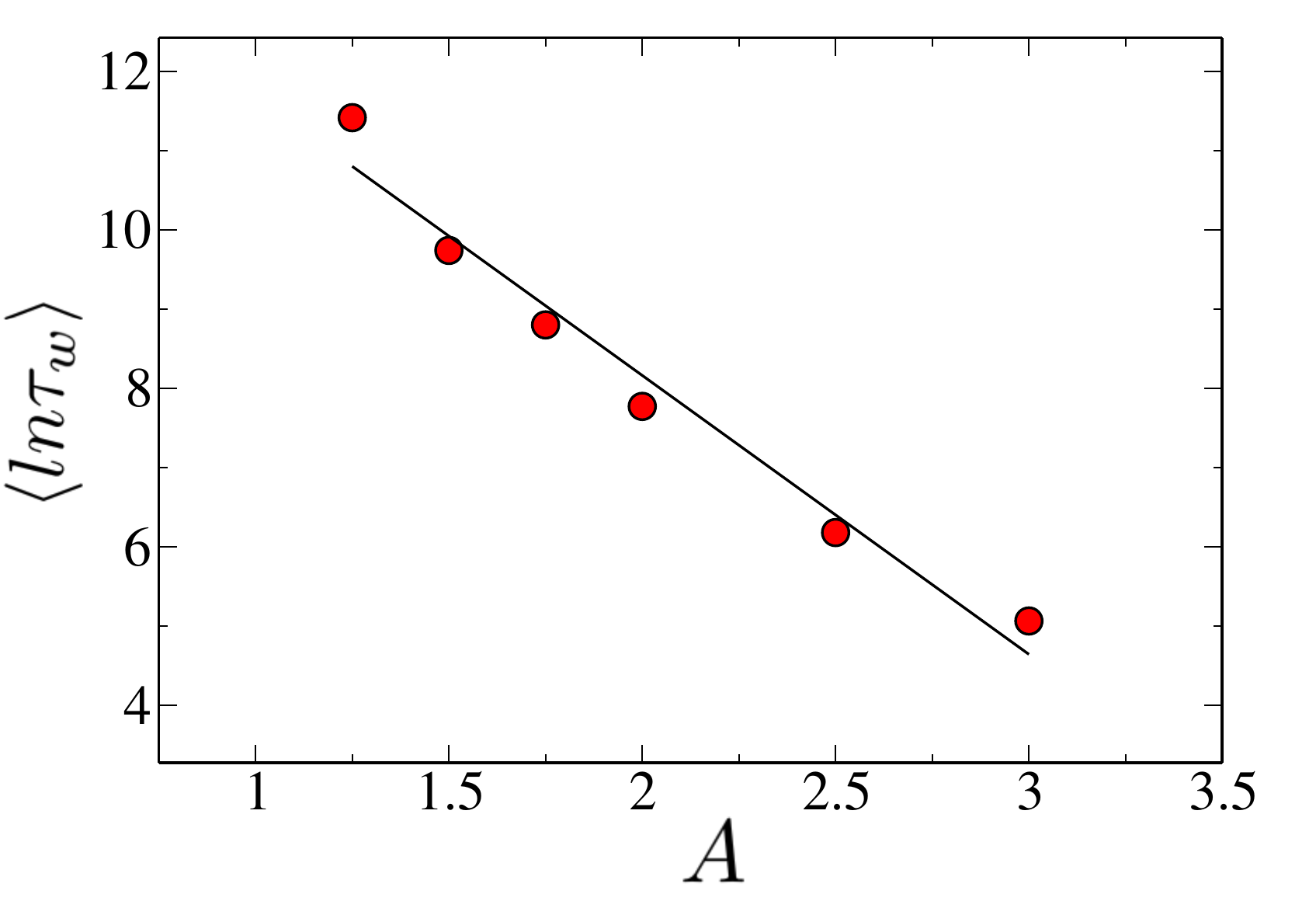}
\caption{Average log of waiting time as a function of aspect ratio for $N = 1000, T = 0.10, \textbf{F}_{app} = 13.0$.}
\label{fig6}
\end{figure}

Next we investigate the dependence on the aspect ratio $A$ for  fixed  $F_{app}$, $T$ and $N$. Once again we find a simple dependence,  
\begin{equation}
\langle \ln \tau_w \left( A \right) \rangle \approx C_2-D_2 A \ .
\label{eqn3}
\end{equation}
The data is shown in Fig.~\ref{fig6}.   
 %%%%%%%%%%%%%%%%%%%%%%%%%%%%%%%%%%%%%%%%%%%%%%%%%%%%%%%%%%%%% 
%%%%%%%%%%%%%%%%%%%%%%%%%%%%%%%%%%%%%%%%%%%%%%%%%%%%%%%%%%%%%%
The coefficient $D_2$ depends on the parameters held fixed in Eq.~(\ref{eqn3}). 
%%%%%%%%%%%%%%%%%%%%%%%%%%%%%%%%%%%%%%%%%%%%%%%%%%%%%%%%%%%%%%%%%%%%%%%%
\begin{figure}
	\hskip 0.3 cm
\includegraphics[scale = 0.4]{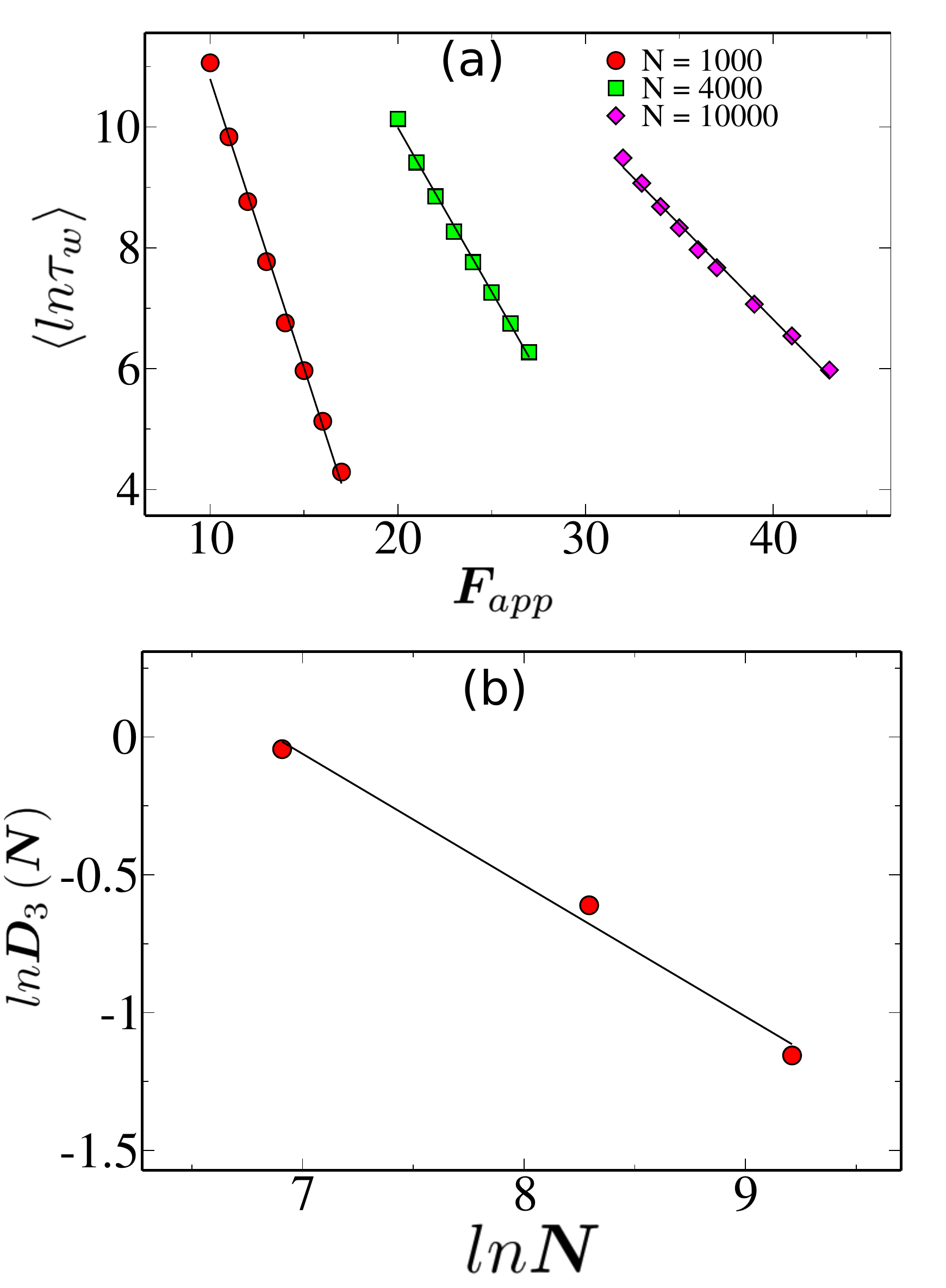}
\caption{Panel (a): Average log of waiting time as a function of $\B F_{app}$ for different system sizes. Panel (b): Variation of the slope, $D_3(T)$, as a function of system size.}
\label{fig7}
\end{figure} 

\subsection{Dependence on system size $N$}

Similarly to the the two previous subsections, we can examine the dependence of the log of time to failure
on the system size. Varying the system size with the other parameters held fixed we find the relation
\begin{equation}
	\langle\ln \tau_{w}\left(F_{app}, N \right)\rangle \approx C_3 - D_3 (N) {F}_{app} \ .
	\label{def3}
\end{equation}
The data to support this law and the dependence of $D_3(N)$ on $N$, (which is consistent with a power law)
are shown in Fig.~\ref{fig7}.

\subsection{Effect of degree of brittleness}

\begin{figure}
\includegraphics[scale = 0.5]{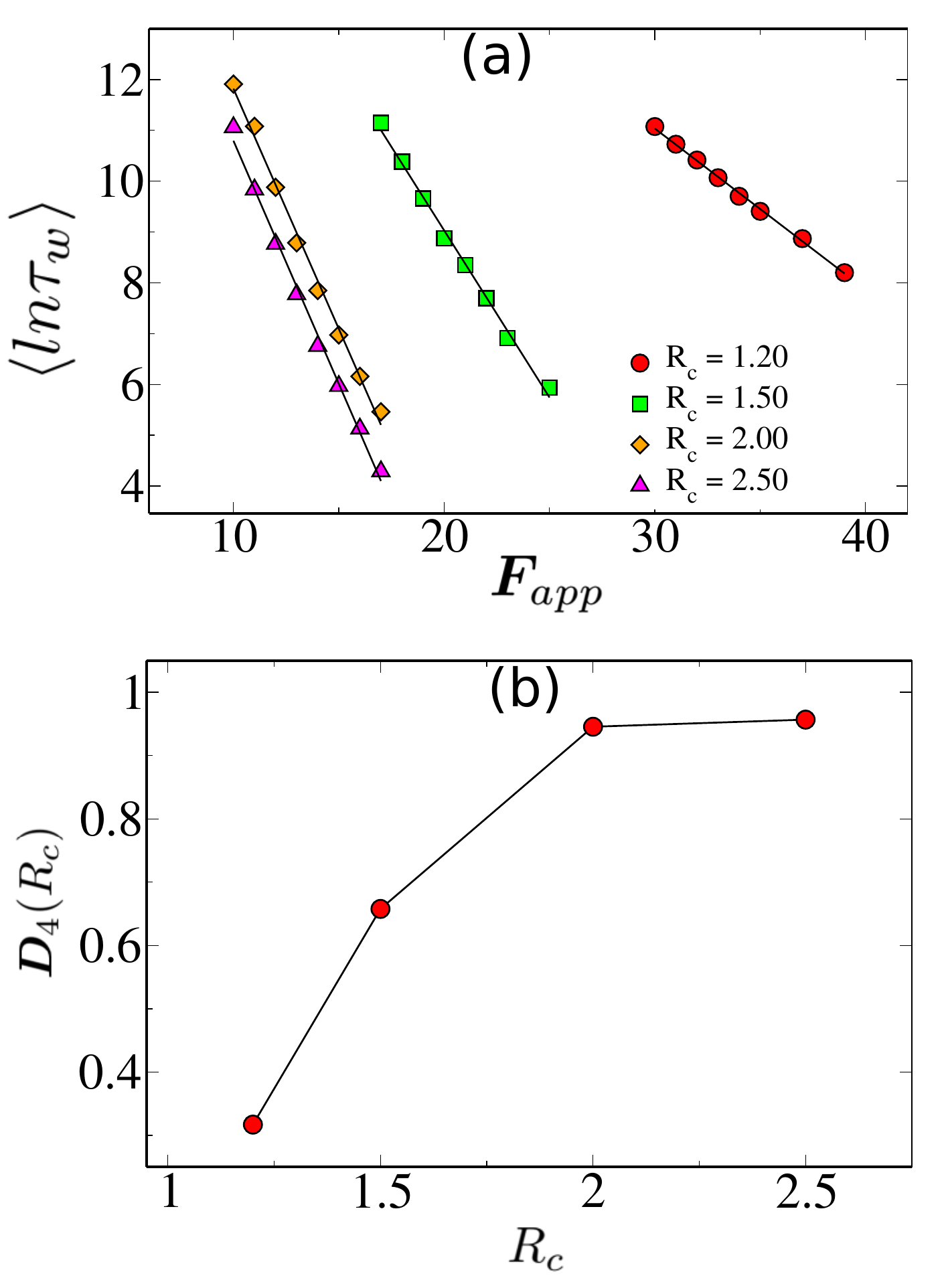}
\caption{Panel (a): Average log of waiting time as a function of $\B F_{app}$ for different interaction range, $R_c$. Panel (b): Variation $D_4$ a function of $R_c$.}
\label{fig8}
\end{figure}

As mentioned above, one can tune the brittleness in the system by changing the interaction range in the force law Eq.~(\ref{LJ}). The cutoff length, $R_c$, is varied from 1.20 to a maximum value of 2.50. With decreasing $R_c$ the  system becomes more brittle, modifying the dependence of $ \langle\ln \tau_{w} \rangle$ on  $F_{app}$:
\begin{equation}
	\langle \ln \tau_{w}\left(F_{app}, R_c\right)\rangle \approx C_4 - D_4 (R_c) {F}_{app}\ .
	\label{def4}
\end{equation}
  The effect is similar to varying the system size $N$, see Fig.~\ref{fig8}. Note that unlike previous case, the coefficient in the exponent does not appear to follow a power law. 
%%%%%%%%%%%%%%%%%%%%%%%%%%%%%%%%%%%%%%%%%%%%%%%%%%%%%%
%%%%%%%%%%%%%%%%%%%%%%%%%%%%%%%%%%%%%%%%%%%%%%%%%%%%

\section{Statistics of the time to failure}
\label{stat}

The upshot of the previous section is that the time to failure depends sensitively on all the parameters
in the system. On top of this, even for a fixed set of parameters, different realization can fail with times
to failure varying on at least two orders of magnitude. Our conclusion is therefore that deterministic 
predictions are quite difficult to make. In this section we turn to the probability distribution function of the time-to-failure, and argue that this path offers quite interesting opportunities for probabilistic predictions. For concreteness we focus on the pdf $P({\ln \tau}_w ; F_{app})$ for systems with a given $N$, $T$ and $R_c$. One can
examine  at will other cuts of parameter space in a similar fashion. 

\subsection{Numerical findings}

\begin{figure*}
\includegraphics[scale = 0.62]{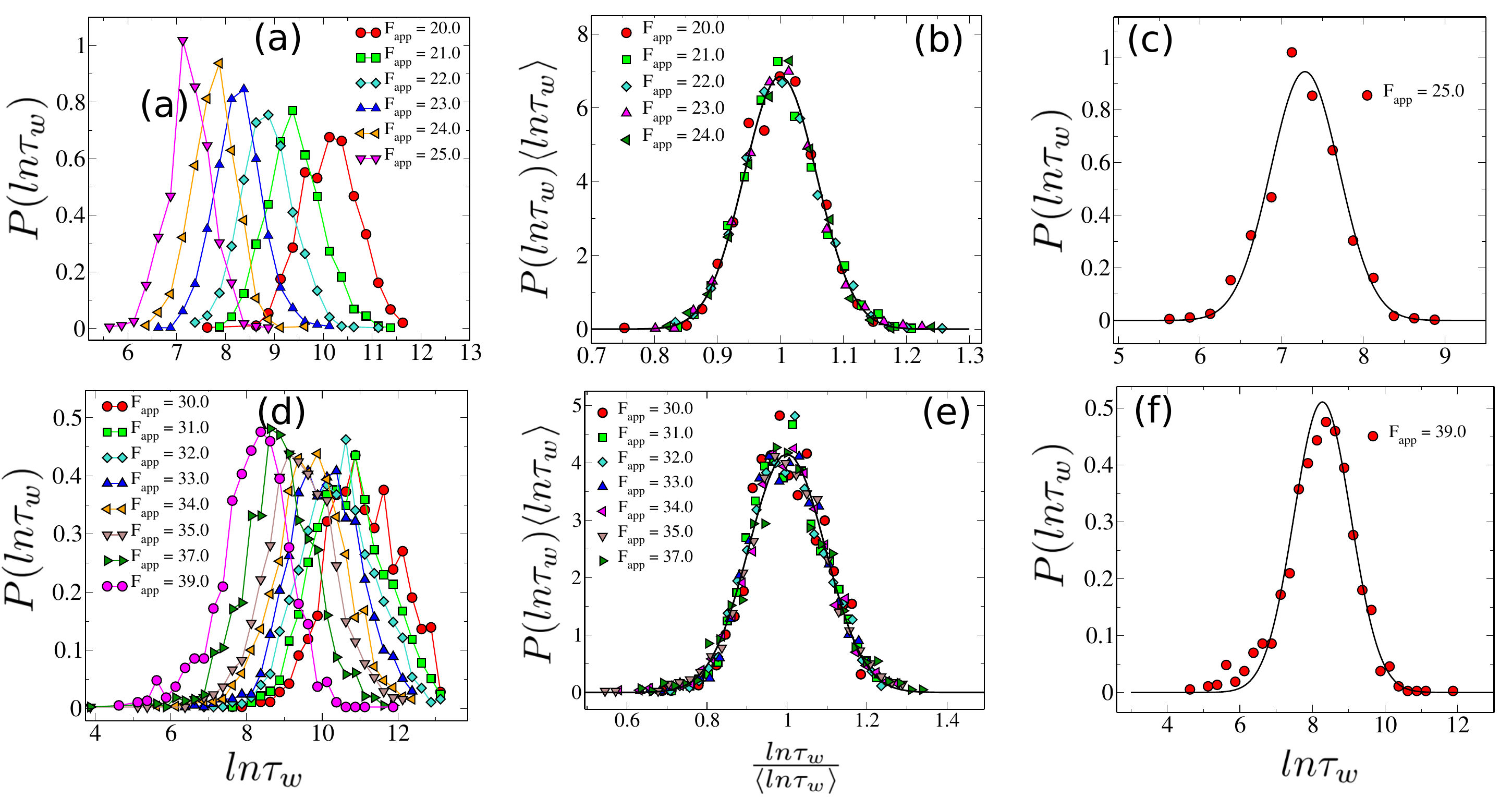}
\caption{Distribution of the $\ln \tau_w$ for various $\B F_{app}$ for both ductile system, panel (a), $N = 4000$, and brittle system in panel (d) for $R_c = 1.2$, $N = 1000$. Note that the qualitative nature of the distribution remains invariant to the degree of brittleness and that the distribution become narrower with increasing system size. Panel (b): Data collapse according to the theoretical prediction Eq.~(\ref{eqn7}) for the ductile case.  Panel (e): Data collapse according to the theoretical prediction Eq.~(\ref{eqn7}) for the brittle case. Panels (c) and (f) demonstrate the prediction of an out-of-sample pdf for ductile and brittle systems respectively. }
\label{fig9}
\end{figure*}

The main finding for the pdf $P({\ln \tau}_w ; F_{app})$ is exhibited in Fig.~\ref{fig9}, panels (a) and (d). The pdf's appear to be log-normal {\em for both brittle and ductile systems}. Of course, for different tensile forces $\B F_{app}$ we find different distributions, but they keep on their form, which is Gaussain in plots of $P(\ln\tau_w)$ vs. $\ln \tau_w$. This suggests that data collapse by scaling should work well, as we show next.
%%%%%%%%%%%%%%%%%%%%%%%%%%%%%%%%%%%%%%%%%%%%%%%%%%%%%%%%%%%%% 
%%%%%%%%%%%%%%%%%%%%%%%%%%%%%%%%%%%%%%%%%%%%%%%%%%%%%%%%%%%%%%%%%%

We find that the simplest scaling ansatz works very well. The pdf $P(\ln \tau_{w}; F_{app})$ can be scaled using the following scaling relation 
\begin{equation}
P\left( {\ln \tau_w}, \langle {\ln \tau}_w \rangle \right) = \langle \ln \tau_{w} \rangle^{-1} g \left( \frac{\ln \tau_{w}}{\langle \ln \tau_{w} \rangle}\right) 
\label{eqn7}
\end{equation}  
The data collapse that follows from this ansatz is presented in Fig.~\ref{fig9} panels (b) and (e) for ductile and brittle systems, with 
\begin{equation}
g(x) = \tilde C \exp [-\hat C(x-1)^2] \ ,
\label{collapsed}
\end{equation}
where $\tilde C$ and $\hat C$ are constants, and $x\equiv \ln \tau_w/\langle \ln \tau_w \rangle$. This data collapse allows statistical predictions as we show next. Note also that Eq.~\ref{collapsed} predicts that the peak of the distribution
of $P\left( {\ln \tau}_w, \langle {\ln \tau}_w \rangle \right)$ occurs when $ {\ln \tau}_w = \langle {\ln \tau}_w \rangle $.
%%%%%%%%%%%%%%%%%%%%%%%%%%%%%%%%%%%%%%%%%%%%
%\begin{figure}
%	\includegraphics[scale = 0.42]{fig10.eps}
%	\caption{Data collapse for the pdf's shown in Fig.~\ref{fig9}. The
%		scaling ansatz Eq.~(\ref{eqn4}) is very well supported. }
%	\label{collapse}
%\end{figure}
%%%%%%%%%%%%%%%%%%%%%%%%%%%%%%%%%%%%%%%%%%%%%%%%%%%%%%%%% 

\subsection{Predictions}

Consider a situation in which data for a limited range of $F_{app}$ is given,
but one needs to estimate the most probable time for failure $\tau_w$ for an out-of-sample value of $F_{app}$. Another important task may be estimating the probability of collapse at times smaller, or even much smaller, than the most probable value of $\tau_w$. Here we show that the analysis described above offer answers to such tasks.  

Imagine then that we have data for the average time for collapse for a range of tensile forces $[F^{min}_{app},F^{max}_{app}]$, and we want to determine the statistics of $\tau_w$ for $F^{oos}_{app}>F^{max}_{app}$ where the superscript "oos" stands for out-of-sample. From the available data we determine the coefficients in Eq.~(\ref{eqn2}), and also those of the collapsed function Eq.~(\ref{collapsed}). Next we can use Eq.~(\ref{eqn2}) to {\em predict} $\langle \ln \tau_w\rangle$ for the out of sample value of the tensile force $F^{oos}_{app}$.
Employing the collapsed pdf Eq.~(\ref{collapsed}) can determine than $P (\ln \tau_w)$, which in turn provides us with full predictability for the statistics of the time-for-failure.

To demonstrate this process consider again the collapsed pdf's in Fig.~\ref{fig9} panels (b) and (e), and pretend that the data for $F_{app}= 25$ and $F_{app}= 39$
did not exist in our sample. Following the procedure outline here we get the predicted distribution shown on the right panels (c) and (f) of Fig.~\ref{fig9}. The agreement with the data is quite satisfactory.

\section{Theory}
\label{theory}

To derive the form of the scaling function for the statistics of $\tau_w$ one cannot avoid delving into the process of material collapse. We find that the ductile case is harder to theorize than the brittle case. Thus in this section we describe the derivation of Eq.~(\ref{eqn7}) for the brittle case, leaving the other for a future endeavor. 

In the brittle case the material collapses due to the growth of damage holes in the material, cf. Fig~\ref{fig2}.
In the brittle case the simulations indicate that a set of damage holes is formed on a rapid time scale, and then the system is dormant for a long time, of the order of $\tau_w$, until a rapid process of increased damage take place until rupture occurs. We therefore need to estimate the time that it takes to break the bonds that form the circumference of the holes present in the system.  We expect to have a distribution of hole areas, but what is important is the circumference of these holes, since it is there that additional bond breaking takes place. Denote then the total length of all the circumferences in a given sample as $\ell$. We assume that the distribution of $\ell$ is normal, and that the distribution of waiting times stems from the sample-to-sample fluctuation in $\ell$, $P_h(\ell)$. We thus write
\begin{equation}
	P_h(\ell) = 
	\frac{1}{\sqrt{2\pi\mu_\ell}} e^{ -\left[\left(\ell -\langle \ell \rangle\right)^2 /2\mu_\ell\right]} \ , \quad \mu_\ell \equiv \langle \ell^2\rangle -\langle \ell\rangle ^2 \ .
	\label{assume}
\end{equation} 

\begin{figure}[!h]
\centering
\includegraphics[scale = 0.40]{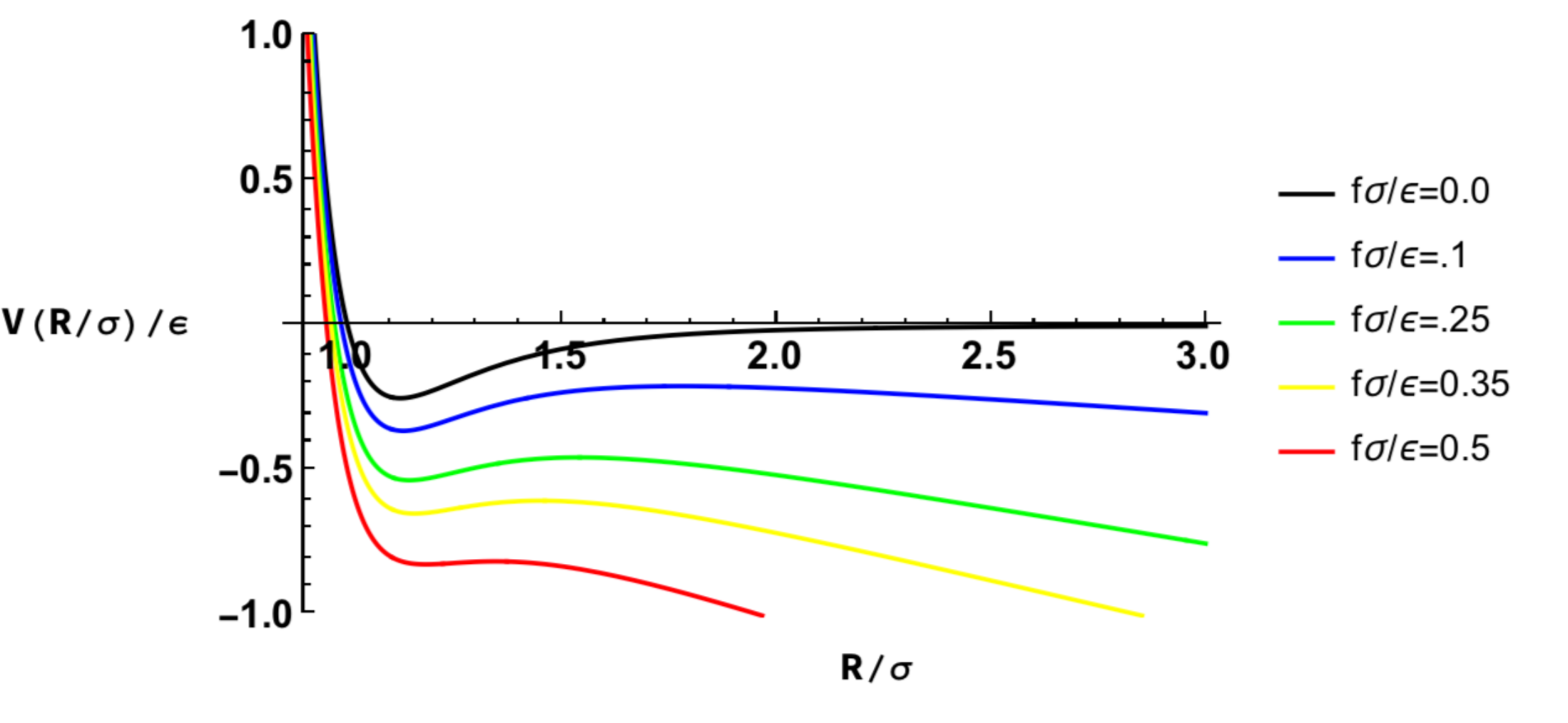}
\caption{Plots of a typical Lennard-Jones Potential under applied strain $f$. Note that for $R>R_{barrier}(f)$ $V(R,f) \rightarrow - \infty$.} 
\label{figvrf}
\end{figure}

Next we estimate the number of bonds associated with the given circumference $\ell$ as $\ell/\sigma$. The energy necessary to be surmounted to break these bonds is $\Delta(f) \ell/\sigma$, where $\Delta(f)$ is the barrier for a {\em single bond} to break under a {\em local} tensile force $f$. Finally we need an expression for this potential barrier. Its actual value in an amorphous strip depends of course on exactly where the bond lies in the strip, and as it will appear as an exponential of the form $\exp{\Delta(f)/T}$ in both thermal averages  and in Kramers' rates for bond breakages \cite{40Kra}, it will fluctuate considerably in its influence. But we can estimate a typical value which depends on the inter-atomic potential used in the simulations. Firstly we estimate the barrier for a simple Lennard-Jones potential (see Fig.~\ref{figvrf}) rather than the more complex form Eq.~(\ref{LJ}). Subsequently we show that the resulting scaling form
for the pdf remains valid also for the potential used, Eq.~(\ref{LJ}).
%%%%%%%%%%%%%%%%%%%%%%%%%%
%%%%%%%%%%%%%%%%%%%%%%%%%%
The strained Lennard Jones potentials  takes the form
\begin{equation}
\label{ref1}
V(R,f) = \epsilon [(R/\sigma)^{-12} -(R/\sigma)^{-6}] - f R.
\end{equation}
Thus the stretched potential is fully specified by three parameters - $\epsilon$ specifies the potential energy scale, $\sigma$ specifies the potential range, while $f$ is the applied force stretching a single bond. 

\begin{figure}[!h]
\centering
\includegraphics[scale = 0.55]{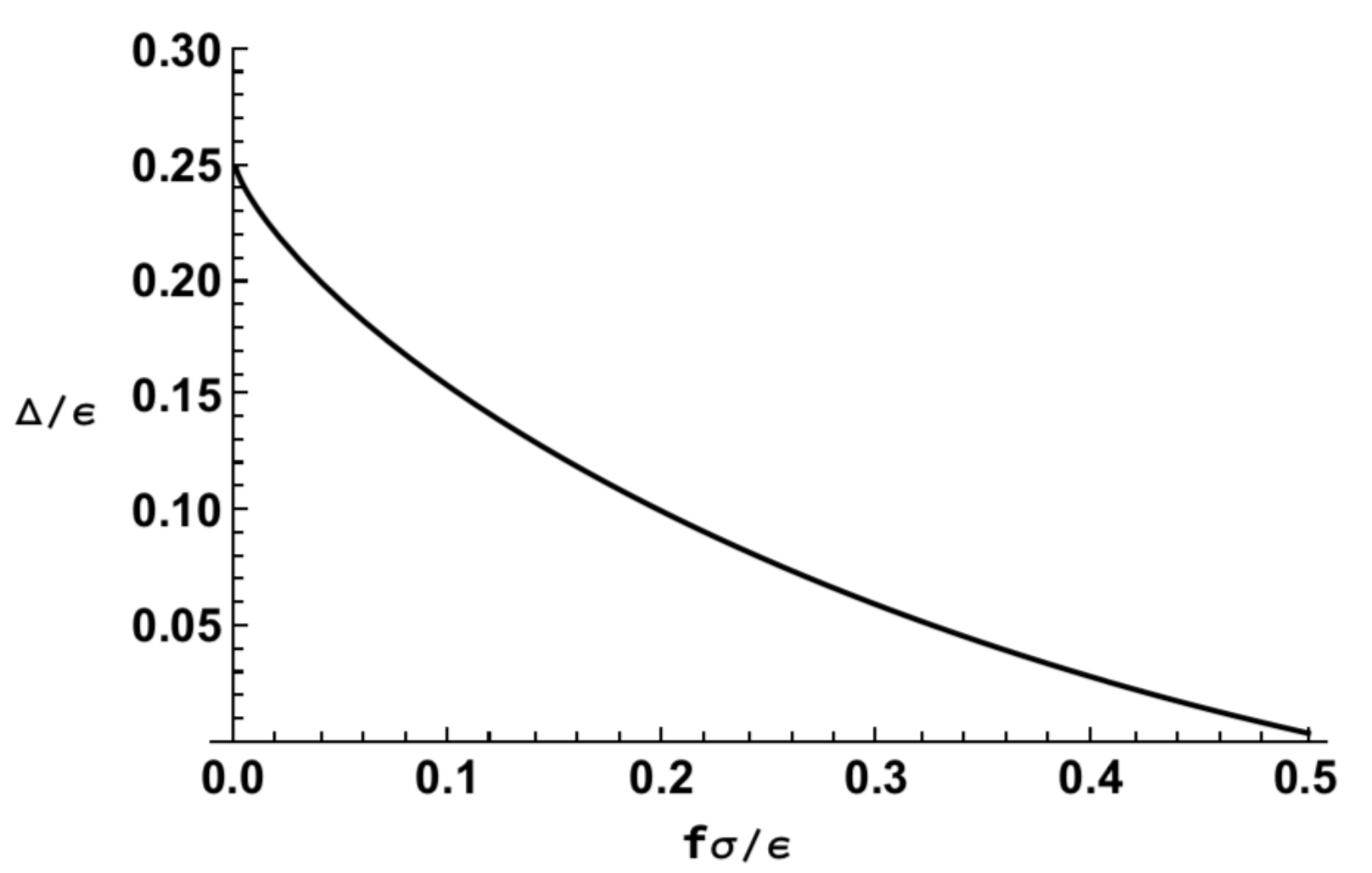}
\caption{The nonlinear scaled barrier height $\Delta (f)/\epsilon$ versus the scaled force $f\sigma/\epsilon$. } 
\label{figljdelta}
\end{figure}

$V(R,f)$ is plotted in Fig.~\ref{figvrf} as a function of $R$ for several values of the applied force $f$. Notice the behaviour of $V(R,f)$  as the strain is increased. Initially an unstable peak in the potential energy barrier of size $\Delta (f)=V(R_{barrier}(f),f) - V(R_{eq}(f),f)$ appears in the potential energy at a distance $R_{barrier}(f)$  from the initial minimum of the unstrained potential. The barrier height $\Delta(f)$ and its position $R_{barrier}(f)$  can be calculated from the pair of equations 
\begin{eqnarray}
\label{bd}
&& \partial V(R,f)/\partial R |_{R = R_{barrier}(f)}  =  0 \nonumber \\ 
&& \partial V(R,f)/\partial R |_{R = R_{eq}(f)}  =  0 \nonumber \\ 
&& \Delta (f)  =  V(R_{barrier}(f),f) - V(R_{eq}(f),f) .
\end{eqnarray}
\noindent Solving Eqs.~(\ref{ref1}) and~(\ref{bd}), we find there is a force $f=f_{max}$ when $\Delta(f_{max})= 0$. The barrier disappears and for $f>f_{max}$  the bond will break. Plotting the nonlinear scaled barrier
 height $\Delta (f)/\epsilon$ versus the scaled force $f\sigma/\epsilon$ in Fig.~\ref{figljdelta} we see that $\Delta(f)$ is a strong function of $f$.
 
%%%%%%%%%%%%%%%%%%%%%%%%%%
%%%%%%%%%%%%%%%%%%%%%%%%%%

Accordingly, we can estimate the time to failure $\tau_w$ as \cite{40Kra}
\begin{equation}
	\ln \tau_w \sim \frac{\ell\Delta(f)}{\sigma T}  \ .
	\label{Kramer}
\end{equation}
The pdf $P(\ln \tau_w)$ can be written as an identity,
\begin{equation}
P(\ln \tau_w) = \int_0^\infty d\ell P_h(\ell)	\delta\left(	\ln \tau_w -  \frac{\ell\Delta(f)}{\sigma T}\right)  \ .
	\end{equation}
Performing the integration and using the $\delta$-function we end up with the log-normal distribution for the waiting times
\begin{eqnarray}
&&	P(\ln \tau_w) = \frac{1}{\sqrt{2\pi\mu_{_{\ln \tau_w}}}} e^{ -\left[\frac{(\ln \tau_w -\langle \ln \tau_w \rangle)^2} {2\mu_{_{\ln \tau_w}}}\right]} \ , \nonumber\\&&\quad \mu_{_{\ln \tau_w}} \equiv \langle (\ln \tau_w)^2\rangle -\langle \ln \tau_w\rangle ^2 \ .
\label{problogtime}
\end{eqnarray}
Finally, we can pull out a factor of $\langle \ln \tau_w\rangle^2$ from the exponential form, and after
some straightforward manipulations bring the pdf to the form used in Fig.~\ref{fig9}. We rewrite Eq.~(\ref{problogtime}) in the form
%\begin{equation}
%P(\ln \tau_w) \langle \ln \tau_w \rangle = \frac{\exp{- \left[\frac{\left(\frac{\ln \tau_w}{\langle\ln \tau_w \rangle} - 1]^2\right)}{2\left(\frac{\langle \ln\tau_w^2 \rangle}{\langle\ln \tau_w\rangle^2} - 1\right)^2}\right]}}{\sqrt{2\pi\left(\frac{\langle \ln \tau _w^2 \rangle}{\langle \ln \tau_w\rangle^2} -1 \right)}}
%\end{equation}
\begin{equation}
P(\ln \tau_w) \langle \ln \tau_w \rangle = \frac{\exp{- \left[\frac{\left(x - 1]^2\right)}{2\mu'_{_{\ln \tau_w}}}\right]}}{\sqrt{2\pi \mu'_{_{\ln \tau_w}}}}
\end{equation}
With $x = \frac{\ln \tau_w}{\langle \ln \tau_w \rangle}$ and $\mu'_{_{\ln \tau_w}} = \frac{\mu_{_{\ln \tau_w}}}{\langle \ln \tau_w\rangle^2}$. We thus derived Eq.~(\ref{collapsed}), with the constants expressed as
\begin{equation}
\bar C= \frac{1}{\sqrt{2\pi\mu'_{_{\ln \tau_w}}}} \, \quad \hat C=\frac{1}{2\pi\mu'_{_{\ln \tau_w}}}	 \ .
	\end{equation}

The data collapse is presented in Fig.~\ref{fig9} panel (e). The black solid line is the scaling function $g(x) = 4.16e^{-54.5837\left(x-1\right)^2}$. The quality of the fit validate our theory for the $P(\ln \tau_w)$ for the brittle system.

Presently we generalize this result for the simulated case of a ternary mixture. The agreement of the scaling theory and the actual simulations seem much better than a theory based on a single LJ potential should allow (see Fig.~\ref{fig9}). We now show that this is not a coincidence because instead of Eq.~(\ref{Kramer}) we can use
\begin{equation}
	\ln \tau_w \sim \frac{\ell \langle \Delta(f)\rangle }{\langle \sigma \rangle T}  \ ,
	\label{Kramer2}
\end{equation}
to estimate the waiting time,  where 
\begin{eqnarray}
	\label{renorm}
	\langle \Delta(f)\rangle & = & \sum_{\alpha\beta}p_{\alpha}p_{\beta} \Delta_{\alpha\beta}(f) \nonumber \\ 
	\langle \sigma \rangle & = & \sum_{\alpha\beta}p_{\alpha}p_{\beta} \sigma_{\alpha\beta} 
\end{eqnarray}
with $p_{\alpha}$ being the fraction of particles of type $\alpha$ in the simulation, while $ \Delta_{\alpha\beta}(f)$ is the barrier that needs to overcome a potential barrier in a bond made from atoms $\alpha$ and $\beta$, while $ \sigma_{\alpha\beta}$ is the radius of the bond. In our simulations there are three types of atoms $\alpha,\beta = A,B,C$, and energy and length scales $\epsilon_{\alpha\beta}$ and $\sigma_{\alpha\beta}$, given below Eq.~(\ref{LJ}).
Note that Eq.~(\ref{Kramer2}) has exactly the same form as for a single Lenard-Jones potential, thus accounting for the fit between theory and simulation shown in Fig.~(\ref{fig9}).

To derive Eqs.~(\ref{Kramer2}) and~(\ref{renorm}) we proceed as follows. We first define $n_{\alpha\beta}(\ell)$ as the number of bonds of type $\alpha\beta$ in an interface of length $\ell$. Now as $n_{\alpha\beta}(\ell) =  p_{\alpha}p_{\beta}\ell/\langle \sigma \rangle$ where $\langle \sigma \rangle = \sum_{\alpha\beta}p_{\alpha}p_{\beta} \sigma_{\alpha\beta}$ is the average radius of a bond in the interface, we can write the total energy required to create this interface as
\begin{equation}
	E(\ell,f)  = \sum_{\alpha\beta}n_{\alpha\beta}(\ell)  \Delta_{\alpha\beta}(f) = \frac{ \ell \langle \Delta(f)\rangle}{\langle \sigma \rangle} ,
	\label{energy}
\end{equation}
which leads to Eq.~(\ref{Kramer2}) together with Eq.~(\ref{renorm}).

\section{Summary and discussion}
\label{summary}

The upshot of this paper is that the details of the creep process depend on many parameters as well as on microscopic interactions, as has been found in many experiments and simulations. Nevertheless, the statistics of $\tau_w$,
as well as the average logarithm time-for-failure $\langle {\ln \tau}_w\rangle$, exhibit relatively simple dependence on the parameters. Moreover, the pdf of $\tau_w$ has log-normal form, allowing us to data collapse the distributions for both the brittle and the ductile cases. This data collapse, together with the simple linear dependence
of $\langle \ln \tau_w\rangle$ on the various parameters, opens up out-of-sample predictions for the pdf of $\ln \tau_w$. Theoretically we could relate, for the brittle case, the
log-normal distribution of $\tau_w$ to the log-normal (assumed) distributions of the circumference of the holes formed at early times in brittle materials. An analog relation of the ductile case is still lacking, and is deferred to future research.

It would be interesting and exciting to test the applicability of the present approach to material science and engineering applications. Hoping that the novel findings of the log-normal statistics holds firm, this opens up an important path for predicting the safety of systems under constant external stress. Using measurements at laboratory accessible parameters, estimates of the danger of creep failure at unmeasured values of stress could be vastly improved.  

{\bf acknowledgments}: This work has been supported in part by the ISF grant \#3492/21, and the Minerva Center for Aging at the Weizmann Institute.

\bibliography{ALL}
\end{document}